\documentclass[12pt,a4paper,superscriptaddress,floatfix]{article}

\advance\voffset by -2.0cm \advance\hoffset by -1.25cm
\usepackage{amsmath}
\usepackage{amssymb}
\usepackage{amsthm}
\usepackage{amsfonts}

\usepackage{mathrsfs}

\numberwithin{equation}{section}

\theoremstyle{definition}

\newcommand{\RR}{\mathbb{R}} 
\newcommand{\NN}{\mathbb{N}} 

\DeclareMathOperator{\nn}{{}^{{}_{\rm x}}_{{}^{\rm x}}}

\newcommand{\D}{\mathcal{D}} 

\newcommand{\be}{\begin{equation}}
\newcommand{\ee}{\end{equation}}

\def\1{\frak 1}
\def\2{\frak 2}
\def\3{\frak 3}

\newlength{\oldcolsep}

\begin{document}

\title{Dual representation for the generating functional of the Feynman path-integral}
\author{Marco Matone}\date{}

\maketitle

\begin{center} Dipartimento di Fisica e Astronomia ``G. Galilei'' \\
 Istituto
Nazionale di Fisica Nucleare \\
Universit\`a di Padova, Via Marzolo, 8-35131 Padova,
Italy\end{center}

\bigskip

\begin{abstract}
\noindent The generating functional for scalar theories admits a representation which is dual with respect to the one introduced by Schwinger,
interchanging the role of the free and interacting terms. It maps $\int V(\delta_J)$ and $J\Delta J$ to $\delta_{\phi_c}\Delta\delta_{\phi_c}$ and
$\int V(\phi_c)$, respectively, with $\phi_c=\int J\Delta$ and $\Delta$ the Feynman propagator. Comparing the Schwinger representation with its dual version one gets
a little known relation
that we prove to be a particular case of a more general operatorial relation.  We then derive a new representation
of the generating functional $T[\phi_c]=W[J]$ expressed in terms of covariant derivatives acting on 1
$$
T[\phi_c] = {N\over N_0} \exp(-U_0[\phi_c])\exp\Big(-\int V({\cal D}_{\phi_c}^-)\Big) \cdot 1
$$
where ${\cal D}_{\phi}^{\pm}(x)=\mp\Delta{\delta\over\delta\phi}(x)+\phi(x)$.
The dual representation, which is deeply related to the Hermite polynomials, is the key to express
the generating functional associated to a sum of potentials in terms of factorized generating functionals. This is applied to renormalization, leading to a factorization of the counterterms of the interaction.
We investigate the structure of the functional generator for normal ordered potentials and derive an infinite set of relations in the case
of the potential ${\lambda\over n!}:\phi^n:$. Such relations are explicitly
derived by using the Fa\`a di Bruno formula. This also yields the explicit expression of the generating functional of connected Green's functions.

\end{abstract}

\newpage

\section{Introduction and summary}

The path-integral is a basic tool in several research fields, quantum mechanics, quantum field theory, statistical mechanics etc. The original idea is due
to Dirac who  observed a deep analogy between the Hamilton-Jacobi theory and the quantum transition amplitudes, proposing the relation \cite{Dirac:1933xn}
\be
\langle q,t| Q,T \rangle \sim e^{-{i\over\hbar}\int_T^t dt L} \ .
\label{Dirac}\ee
Subsequently, Feynman developed Dirac's idea, starting from an infinitesimal version of (\ref{Dirac}), another key step toward the path-integral formulation \cite{Feynman}.

\noindent
A relevant progress in the path-integral approach is due to Schwinger who expressed the generating functional in the form
\be
W[J]={N\over N_0}\exp\Big(-\int V\Big({\delta\over\delta J}\Big)\Big)\exp(-Z_0[J]) \ ,
\label{diventa}\ee
where $V$ is the potential, $J$ the external source and $\exp(-Z_0[J])$ the generating functional of the free theory.
It turns out that the Schwinger representation admits the dual representation
\be
W[J]={N\over N_0}\exp(-Z_0[J])\exp\Big({1\over2}{\delta\over\delta J}\Delta^{-1}{\delta\over\delta J}\Big)
\exp\Big[-\int d^Dx V\Big(\int d^Dz J(z)\Delta(z-x)\Big)\Big] \ ,
\label{rdue}\ee
with $\Delta(y-x)$ the Feynman propagator.
Such a dual representation leads to consider the field
\be
\phi_c(x)=\int d^DyJ(y)\Delta(y-x) \ ,
\ee
rather than $J$ and then defining $T[\phi_c]=W[J]$.

\noindent As we will see, this leads to represent the path-integral operator as an operator acting by functional derivatives. Namely, for any functional
$F[\phi]$, it holds
\be
N_0 \int D\phi \exp\Big(-{1\over 2} \phi \Delta \phi\Big) F[\phi] = \exp\Big({1\over 2}{\delta\over\delta \chi} \Delta {\delta\over\delta \chi} \Big) F[\chi] |_{\chi=0} \ .
\label{pathderivative}\ee
This is a consequence of the general relation
\be
\langle0|TF[\hat\phi+g]|0\rangle =\exp\Big({1\over2}{\delta\over \delta g} \Delta{\delta\over\delta g}\Big)F[g] \ ,
\ee
derived in Sec. \ref{sec-due}. We then will derive a new representation of the generating functional that simplifies considerably
the computations. Namely, in Sec. \ref{covariant}, we will see that $T[\phi_c]$ can be expressed in terms of covariant derivatives acting on 1, that is
\be
T[\phi_c] = {N\over N_0} \exp(-U_0[\phi_c])\exp\Big(-\int V({\cal D}_{\phi_c}^-)\Big) \cdot 1 \ ,
\label{questA}\ee
where
\be
{\cal D}_{\phi}^{\pm}(x)=\mp\Delta{\delta\over\delta\phi}(x)+\phi(x) \ .
\ee

\noindent
We will derive the little-known representation (\ref{rdue}), reported in Fried's book \cite{FriedBook}, in two different ways. In Sec. \ref{sec-due} Eq.(\ref{rdue}) is derived using the path-integral
and the operator formalism.
Then, in Sec. \ref{sec:dual}, we will derive a more general operatorial relation, Eq.(\ref{generalissima}),
that, in turn, is the functional analog of an operatorial relation satisfied by the Hermite polynomials.
In particular,
Eq.(\ref{generalissima}) implies the
operatorial identity
\begin{align}
 &\exp\Big(-\int V({\delta_J})\Big)\exp\Big({1\over2} J \Delta J\Big) \cr
 & = \exp\Big(-{1\over2}\phi_c\Delta^{-1}\phi_c\Big)\Big[\exp\Big({1\over2}{\delta_{\phi_c}}\Delta\delta_{\phi_c}\Big)
 \exp\Big(-\int V(\phi_c)\Big)\Big]
  [\Delta \delta_{\phi_c}]\exp\Big(\chi \Delta^{-1}  \phi_c\Big)|_{\chi= \phi_c}   \ , \cr
\label{outstanding}\end{align}
that, applied to a constant, reproduces the relation implied by the identification of (\ref{diventa}) with (\ref{rdue}).

\noindent We note that the operator $\exp({1\over2}{\delta_J}\Delta^{-1}{\delta_J})$
appears also in the context of renormalization, see e.g.
\cite{Keller:1991bz}\cite{Salmhofer:1998nk} and references therein.

\noindent
As will be clear from the
investigation, the dual representation is deeply related to the Hermite polynomials which appear in several contexts. For example, in Sec. \ref{sec-due}, it is shown that
the dual representation is the functional generalization of the Weierstrass representation of the Hermite polynomials. Furthermore, in subsec. \ref{SDSD},
it is shown how the Hermite polynomials arise in the Schwinger-Dyson equation expressed in terms of the dual representation of the generating functional. Another result concerns
the following expression of the Schwinger-Dyson equation for the normal ordered potential $:V:$
\be
\Big[{\delta\over\delta\phi_c(x)}+e^{2U_0[\phi_c]} \int{\delta V\over\delta \phi(x)}\Big(\Delta {\delta\over \delta \phi_c}\Big)e^{\chi \Delta^{-1}\phi_c}|_{\chi=\phi_c}\Big]e^{{1\over2}{\delta\over \delta \phi_c} \Delta{\delta\over\delta \phi_c}}
e^{-\int :V:  (\phi_c)}=0 \ .
\label{SDnewstAAA}\ee
Even such an equation, anticipated in subsec. \ref{SDSD}, follows by the operatorial relation (\ref{generalissima}). In Sec. \ref{sec-due} we will also show that the dual representation is
naturally related to the $S$-operator. In particular, it turns out that such an operator is proportional to the normal ordered version of the dual generating functional with $\phi_c$ replaced
by the operator $\hat\phi$, that is
\be
S[\hat \phi]=N^{-1}: T[\hat\phi] : \ .
\ee

\noindent The functional structure of $T[\phi_c]$ suggests considering the following problem investigated in Sec. \ref{sec-factor}. Given a potential corresponding to a summation of potentials,
\be
V(\phi)=\sum_{k=1}^n V_k(\phi) \ ,
\ee
express the full generating functional $T[\phi_c]$ associated to $V$ in terms of the generating functionals $T_k[\phi_c]$ associated to the potential $V_k$, $k=1,\ldots,n$.
It turns out that, for $n=2$,
\be
T[\phi_c] =  \exp(\delta_{\phi_{c_1}}\Delta\delta_{\phi_{c_2}})\exp\theta(1,2)T_1[\phi_{c_1}]T_2[\phi_{c_2}]|_{\phi_{c_1}=\phi_{c_2}=\phi_{c}} \ ,
\ee
where
\be
\theta(1,2)=-U_0[\phi_{c}]+U_0[\phi_{c_1}]+U_0[\phi_{c_2}] \ ,
\ee
with $U_0[\phi_c]=Z_0[J]$, and the normalization constants absorbed in $T[\phi_c]$ and $T_k[\phi_c]$.
Such an investigation, that can be extended to the case of fermions and gauge fields, has several applications. Here we consider the case in which the potential is given by the summation of the original one and the
one coming from the counterterms. The outcome is the following relation (see Sec. \ref{sec-factor} for the notation)
\be
T_{ren}[\phi_c] = \exp(\delta_{\phi_{c_1}}\hat \Delta\delta_{\phi_{c_2}})\exp\hat\theta(1,2)\hat T[\phi_{c_1}]\hat T_{ct}[\phi_{c_2}]|_{\phi_{c_1}=\phi_{c_2}=\phi_{c}}  \ .
\ee
We also show that imposing the associativity condition in the case $n=3$ one gets the relation
\be
\exp(\delta_{\phi_{c_{A}}} \Delta\delta_{\phi_{c_B}})\exp\theta(A,B)\Big(T_{12}[\phi_{c_{A}}]T_3[\phi_{c_B}]-T_{1}[\phi_{c_{A}}]T_{23}[\phi_{c_B}]\Big)|_{\phi_{c_{A}}=\phi_{c_B}=\phi_{c}}=0 \ ,
\ee
where $T_{jk}$ denotes the generating functional associated to the potential $V_j+V_k$.

\noindent
Sec. \ref{sec:dual} is devoted to the derivation of the operatorial relation from which follows, as a particular case, Eq.(\ref{outstanding}), which, in turn, implies the identification of
(\ref{diventa}) with (\ref{rdue}). We then derive, by a different method, the analogous relation in the case of the Hermite polynomials.

\noindent
In Sec. \ref{sec-five} we consider the functional generator
in the case of normal ordered potentials. There is a considerable simplification of the general expression since the normal ordering precisely cancels the action of the operator
$\exp({1\over2}\delta_{\phi_c}\Delta\delta_{\phi_c})$ acting on each single potential coming from the expansion. In particular, the functional generator of the connected Green's functions
$U[\phi_c]=-\ln T[\phi_c]$ turns out to be
\be
U[\phi_c]=\ln {N_0\over N}+U_0[\phi_c]+\sum_{p=1}^\infty {(-1)^{p+1}\over p!}\prod_{j>k}^p e^{\D_{jk}}\prod_{i=1}^p \int V(\phi_{c_i})|_{c, \phi_{c_1},\ldots,\phi_{c_p}=\phi_c} \ ,
\label{hgft}\ee
where
\be
 \D_{jk}={\delta\over \delta \phi_{c_j}} \Delta{\delta\over\delta \phi_{c_k}}  \ ,
\ee
$j,k\in \NN_+$. Comparing the expression Eq.(\ref{hgft}) with the explicit expression of the generating functional
for the potentials ${\lambda\over n!}:\phi^n:$, obtained in \cite{Matone:2015nxa}, leads to an identity that implies an infinite set of relations. Such relations are explicitly
derived by using the Fa\`a di Bruno formula giving the chain rules for higher order derivatives. In particular, we will derive the following explicit expression
for the action of the operators  $\exp\D_{jk}$
\be
\prod_{j>k}^p e^{\D_{jk}}\prod_{i=1}^p \int d^Dz_l \phi_{ci}^n(z_l)|_{c, \phi_{c_1},\ldots,\phi_{c_p}=\phi_c}= (n!)^p p!\sum_{l=1}^p (-1)^{l+1}\sum_{j_1+\cdots+j_l=p} {h_{n,j_1}\over j_1!}\cdots {h_{n,j_l}\over j_l!} \ ,
\ee
where the $h_{n,k}$ are multinomials of integrated powers of $\phi_c$. By (\ref{hgft}) this also yields the explicit expression of $U[\phi_c]$.
In subsec. \ref{dualphi}
we show a duality between the field $\phi$ and the source $\phi_c$. This suggests promoting $\phi_c$ to a dynamical field, leading
to a potential $V(\phi+\phi_c)$. It turns out that the model is equivalent to the one of a single field $\phi_c$ with $U[\phi_c]$ playing the role
of potential. Iterating the construction leads to a rescaling of the kinetic term. This seems a possible alternative with respect to the renormalization procedure
where the rescaling of the kinetic term is obtained by rescaling the field itself.

\noindent
In Sec. \ref{covariant} we derive the new representation (\ref{questA}) of the generating functional. We also show how such a representation
simplifies the calculations. In particular, the action of the normal ordering operator on a product of functionals simplifies considerable, namely
\be
\exp\big(-{{1\over2}  \delta_{\phi}\Delta\delta_{\phi}}\big)F[\phi]G[\phi] = F[{\cal D}^+_{\phi}]G[{\cal D}^+_{\phi}]\cdot 1 \ .
\ee

\section{Dual representation of the generating functional}\label{sec-due}

For any even function or distribution $h$, and any functions or operators $f$ and $g$, set
\be
f h g = \int d^D x\int d^D y f(x)h(x-y)g(y) \ ,
\ee
and
\be
{\delta\over\delta f}h{\delta\over\delta g}=\int d^D x\int d^D y {\delta\over\delta f(x)}h(x-y){\delta\over\delta g(y)} \ .
\ee
We will also use the notation
\be
hg(x)=\int d^Dyh(x-y)g(y) \ ,
\ee
and
\be
h {\delta\over \delta g}(x)=\int d^Dyh(x-y) {\delta\over \delta g(y)} \ .
\ee
Consider the Feynman propagator
\be
\Delta(x-y)=\int {d^Dp \over (2\pi)^D} {e^{ip(x-y)}\over p^2+m^2} \ ,
\ee
and its inverse
\be
\Delta^{-1}(y-x)=(-\partial^2+m^2)\delta(y-x)=\int {d^Dp\over (2\pi)^D}(p^2+m^2) e^{ip(y-x)} \ .
\ee
Denote by $Z_0[J]$ the functional generator of the free connected Green's function, that is
\be
Z_0[J]=-{1\over2}J\Delta J \ .
\label{zzero}\ee
We will focus our investigation on a scalar field in the $D$-dimensional Euclidean space. The corresponding generating functional is
\be
W[J] = N\int D\phi \exp\Big(-S[\phi]+\int J\phi\Big) \ ,
\label{primordiale}\ee
where $J$ is the external source,
\be
S[\phi] = \int d^Dx \Big({1\over2}\partial_\mu\phi\partial_\mu\phi + {1\over2}m^2\phi^2 +V(\phi)\Big) \ ,
\ee
is the scalar action with potential $V(\phi)$, and
$N=1/\int D\phi \exp(-S)$ the normalization constant. We denote by $S_0[\phi]$ the free action.

\noindent To fix the normalization constant in (\ref{diventa}), note that
\be
\exp\Big(-\int V\Big({\delta\over\delta J}\Big)\Big) \exp(-Z_0[J])={\int D\phi \exp(-S[\phi]+\int J\phi)\over  \int D\phi \exp(-S_0[\phi])} \ ,
\label{formally}\ee
has the correct normalization. This follows by two checks. The first is to shift $V$ by a $L^1(\RR^D)$ function.
The other one is to set $V=0$ and $J=0$, giving 1 in both sides.
It follows that the normalization of the Schwinger representation is the one in (\ref{diventa}) with
$N_0=1/\int D\phi \exp(-S_0)$. Such a normalization is usually omitted in extracting the term $\exp(-\int V(\delta_J))$ from the path
integral, identifying $\int D\phi \exp (-S_0[\phi]+\int J\phi)$, rather than $N_0\int D\phi \exp (-S_0[\phi]+\int J\phi)$, with $\exp(-Z_0[J])$.

\noindent
The correspondence between the Schwinger representation and the operator formalism follows by
\be
W[J]={\langle \Omega|\Omega\rangle_J\over\langle\Omega|\Omega\rangle} \ ,
\label{norma}\ee
where $|\Omega\rangle$ is the vacuum of the interacting theory and $\langle \Omega|\Omega\rangle_J$ denotes the vacuum-vacuum amplitude in the
presence of the external source $J$. In particular, denoting by $|0\rangle$ the free vacuum, normalized by $\langle0|0\rangle=1$, we have
$\langle 0|0\rangle_J=\langle 0| T\exp(\int J\hat\phi) | 0\rangle$. Therefore,
\be
W[J] = \exp(-Z[J]) = {N}\langle 0|T \exp\Big[{\int (-V(\hat\phi)+J\hat\phi)}\Big]|0\rangle\ ,
\label{anche}\ee

\noindent Eq.(\ref{rdue}) has been observed in the case of the exponential interaction, considered as master potential, in \cite{Matone:2015nxa}.
We will prove (\ref{rdue}) using the Euclidean time-ordering, defined by the analytic continuation from
the one in Minkowski space.
Excellent references for the analytic continuation and related issues in the axiomatic approach
include \cite{Streater:1989vi}-\cite{Eckmann:1979vq}.

\noindent
Consider the field
\be
\phi_c(x)=\int d^Dy J(y)\Delta(y-x) \ ,
\ee
satisfying the classical equation of motion with $V=0$ and external source $J$
\be
(-\partial^2+m^2)\phi_c=J \ .
\label{laclassica}\ee
To prove (\ref{rdue}), we first consider the shift
\be
\phi=\phi\,'+\phi_c \ ,
\ee
in the generating functional of the interacting
theory (\ref{anche}), rather than, as usual, on the free one. Dropping the prime in $\phi\,'$, yields
\be
W[J]={N}\exp(-Z_0[J])\langle 0 | T \exp\Big(-\int  V(\hat\phi+\phi_c)\Big)|0\rangle \ ,
\label{showsthat}\ee
which is equivalent to
\be
W[J]=N\exp(-Z_0[J])\int D\phi \exp\Big(- {1\over2}\phi\Delta^{-1}\phi-\int V(\phi+\phi_c)\Big)  \ .
\label{showsss}\ee
Let $F[\hat\phi]$ be a functional of the field operator $\hat\phi$. According to Wick's theorem
\be
TF[\hat\phi]= \exp\Big({1\over2}{\delta\over \delta \hat\phi} \Delta{\delta\over\delta \hat\phi}\Big):  F[\hat\phi]:   \ .
\label{WickTh}\ee
Next, note that for any function $g(x)$
\begin{align}
\langle0|TF[\hat\phi+g]|0\rangle &= \langle0|\exp\Big({1\over2}{\delta\over \delta g} \Delta{\delta\over\delta g}\Big):  F[\hat\phi+g] :  |0\rangle \cr
& =\exp\Big({1\over2}{\delta\over \delta g} \Delta{\delta\over\delta g}\Big)\langle0|F[g]|0\rangle
\cr &= \exp\Big({1\over2}{\delta\over \delta g} \Delta{\delta\over\delta g}\Big)F[g] \ ,
\label{normall}\end{align}
where we used $\langle 0|:G[\hat\phi]:|0\rangle=G[0]$,
holding for any functional $G$ of $\hat\phi$.
Eq.(\ref{rdue}) then follows by applying (\ref{normall}) to the vev in (\ref{showsthat}).

\noindent The relation (\ref{normall}) is quite general and can be used to evaluate any correlator. In particular, note that
\be
\langle0|TF[\hat\phi]|0\rangle = \langle0|TF[\hat\phi+\chi]|0\rangle_{\chi=0}=\exp\Big({1\over2}{\delta\over \delta \chi} \Delta{\delta\over\delta \chi}\Big)F[\chi]|_{\chi=0} \ .
\label{fgr}\ee
This implies
that the path-integral operator is equivalent to the action of a functional derivative. This is Eq.(\ref{pathderivative}), namely
\be
{\int D\phi \exp\big(-{1\over 2} \phi \Delta \phi\big) F[\phi]\over \int D\phi \exp\big(-{1\over 2} \phi \Delta \phi\big)}=
\exp\Big({1\over 2}{\delta\over\delta \chi} \Delta {\delta\over\delta \chi} \Big) F[\chi] |_{\chi=0} \ .
\label{pathderivativebisse}\ee
Applying (\ref{fgr}) to
(\ref{anche}), yields
\be
W[J] =  {N\over N_0}   \exp\Big({1\over2}{\delta\over \delta \chi} \Delta{\delta\over\delta \chi}\Big) \exp\Big[{\int (-V(\chi)+J\chi)}\Big]|_{\chi=0} \ .
\label{anchese}\ee
Taking the multiple derivative of $W[J]$ with respect to $J$ or, equivalently, choosing
\be
F[\hat \phi]=\hat\phi(x_1)\ldots\hat\phi(x_N) \exp\Big(-\int V(\hat\phi)\Big) \ ,
\ee
yields
\be
{\langle \Omega |T\hat\phi(x_1)\ldots\hat\phi(x_N)|\Omega\rangle \over \langle \Omega|\Omega\rangle}
= {N\over N_0} \exp\Big({1\over2}{\delta\over \delta \chi} \Delta{\delta\over\delta \chi}\Big) \chi(x_1)\cdots \chi(x_N)\exp\Big({-\int V(\chi)}\Big)|_{\chi=0} \ .
\ee

\noindent
We note that if $TF[\hat\phi]=F[\hat\phi]$, then $\exp(-{1\over2}{\delta_{\hat\phi}} \Delta{\delta_{\hat\phi}})$ is the normal ordering operator. When the argument
of $F$ is not an operator, then $TF[\phi]=F[\phi]$, and the map
\be
F[\phi]\longrightarrow \exp\Big(-{1\over2}{\delta\over \delta \phi} \Delta{\delta\over\delta \phi}\Big)F[\phi] \ ,
\label{rhsof}\ee
is sometimes called Wick transform of $F[\phi]$. Nevertheless, in the following we will call $\exp(-{1\over2}{\delta_\phi} \Delta{\delta_\phi})$
normal ordering operator even in this case, and will also interchange the notation on the right hand side of (\ref{rhsof}) with $:F[\phi]:$.

\subsection{$T[\phi_c]=W[J]$}

The structure of Eq.(\ref{rdue}) suggests considering the generating functional
\be
T[\phi_c]=W[J] \ ,
\label{pdual}\ee
so that by (\ref{diventa})
\be
T[\phi_c]={N\over N_0}\exp\Big({1\over2}\phi_c\Delta^{-1} \phi_c \Big)\exp\Big({1\over2}{\delta\over \delta \phi_c} \Delta{\delta\over\delta \phi_c}\Big) \exp\Big(-\int V(\phi_c)\Big) \ .
\label{quindici}\ee
In the following we will set $U_0[\phi_c]=Z_0[J]$, that is
\be
U_0[\phi_c]=-{1\over2}\phi_c\Delta^{-1} \phi_c \ .
\ee
Eq.(\ref{pdual}) is just the {\it dual relation}
\be
\exp\Big(-{1\over2}J\Delta J\Big) \exp\Big(-\int V\Big({\delta\over \delta J}\Big)\Big)\exp\Big({1\over2}J\Delta J\Big)  = \exp\Big({1\over2}{\delta\over\delta\phi_c}\Delta {\delta\over\delta\phi_c}\Big)
\exp\Big(-\int V(\phi_c)\Big) \ .
\label{daglie}\ee
As shown in Sec. \ref{sec:dual}, such an identity also follows by an operatorial identity once it acts on a constant.
Furthermore, by (\ref{showsss}) and (\ref{quindici}), we have
\be
\exp\Big({1\over2} {\delta_{\phi_c}} \Delta{\delta_{\phi_c}}\Big) \exp\Big(-\int V(\phi_c)\Big)
= {\int D\phi \exp\big(- {1\over2}\phi\Delta^{-1}\phi-\int V(\phi+\phi_c)\big)\over \int D\phi\exp\big(-{1\over2}\phi\Delta^{-1}\phi\big)} \ .
\label{weierstrass}\ee
We will see that this is the functional generalization of the Weierstrass transform.

\noindent
Another property of $T[\phi_c]$ concerns its relation with the effective action
\be
\Gamma[\phi_{cl}]=Z[J]-\int d^D x J(x){\delta Z[J]\over\delta J(x)} \ ,
\ee
where
\be
\phi_{cl}(x)=-{\delta Z[J]\over\delta J(x)} \ .
\ee
Since the map $J\to \phi_c$ is linear, it follows that the Legendre transform of $Z[J]$ is the same of the one of
\be
U[\phi_c]=Z[J] \ ,
\ee
that is
\be
\Sigma[\bar\phi]=U[\phi_c]-\int d^Dx\phi_c(x){\delta U[\phi_c]\over\delta \phi_c(x)}=\Gamma[\phi_{cl}]\ ,
\ee
where
\be
\bar\phi(x)=-{\delta U[\phi_c]\over\delta \phi_c(x)}=\Delta^{-1}\phi_{cl}(x) \ .
\ee

\subsection{$S[\hat\phi]=N^{-1}:T[\hat\phi]:$}

Let us start by showing that the $S$-operator has a simple expression in terms of $T[\hat\phi]$.
To this end note that while the functional derivatives of $T[\phi_c]$  with respect to $J$ yield the
Green functions, deriving with respect to $\phi_c$ gives
\begin{align}
&F^{(k)}(  x_1, \ldots,x_k):={\delta^k T[\phi_c]\over \delta \phi_c(x_1)\ldots\delta \phi_c(x_k)}{\big|_{\phi_c=0}} \cr
&=\int d^Dy_1\ldots\int d^Dy_k \Delta^{-1}(x_1-y_1)\ldots \Delta^{-1}(x_k-y_k) {\langle \Omega| T \phi(y_1)\ldots \phi(y_k)|\Omega\rangle\over \langle\Omega|\Omega\rangle} \ .
\end{align}
Comparing this with the expansion of the $S[\hat\phi]$ operator
\be
S[\hat\phi]=N^{-1}\sum_{k=0}^\infty {1\over k!} \int d^D x_1\ldots\int d^D x_k F^{(k)}(x_1, \ldots, x_k):\hat\phi(y_1)\ldots \hat\phi(y_k):   \ ,
\label{consist}\ee
yields
\be
S[\hat\phi]=N^{-1}:T[\hat\phi]: \ .
\label{questar}\ee
On the other hand, using $\exp({\hat\phi}{\delta_{\chi}})$ that acts by translating $\chi$ by $\hat\phi$,
we have
\be
:T[\hat\phi]:= :\exp\Big({\hat\phi}{\delta\over\delta \chi}\Big): T[\chi]|_{{\chi}=0} \ .
\ee
Replacing $T[\chi]$ by the right hand side of (\ref{quindici}), we get
\begin{align}
S[\hat\phi] & =  {1\over N_0} :\exp\Big({\hat\phi}{\delta\over\delta \chi}\Big): \exp\Big({1\over2}\chi\Delta^{-1}\chi\Big)
 \exp\Big({1\over2}{\delta\over \delta\chi}\Delta{\delta\over\delta\chi}\Big) \exp\Big(-\int V(\chi)\Big)|_{\chi=0} \cr
& =  {1\over N_0} :\exp\Big({1\over2}\hat\phi\Delta^{-1}\hat\phi\Big)\exp\Big({1\over2}{\delta\over \delta\chi}\Delta{\delta\over\delta\chi}\Big)  \exp\Big(-\int V(\hat\phi+\chi)\Big):|_{\chi=0} \ ,
\end{align}
that is
\be
S[\hat\phi] = {1\over N_0} \exp\Big({1\over2}{\delta\over \delta\chi}\Delta{\delta\over\delta\chi}\Big) :\exp\Big({1\over2}\hat\phi\Delta^{-1}\hat\phi-\int V(\hat\phi+\chi)\Big):|_{\chi=0} \ .
\label{suchin}\ee
Using the path-integral representation of $T[\hat\phi]$, it follows by  (\ref{questar}) that such an expression is equivalent to
\be
S[\hat\phi]=\int D\phi \exp(-S[\phi]):\exp (\phi\Delta^{-1}\hat\phi): \ .
\label{suchan}\ee
Then, the representation (\ref{pathderivativebisse}) of the path-integral yields
\be
S[\hat\phi] = {1\over N_0}\exp\Big({1\over2} {\delta\over\delta\chi}\Delta{\delta\over\delta\chi}\Big) \exp\Big(-\int V(\chi)\Big) : \exp(\chi \Delta^{-1}\hat\phi):|_{\chi=0} \ .
\label{suchinvvv}\ee

\subsection{Relation with the Weierstrass transform and the Hermite polynomials}

The Weierstrass transform can be seen as a particular case of (\ref{weierstrass}).
This arises by first considering the Laplace transform of the Gaussian
\be
e^{{t^2\over2}}={1\over \sqrt{2\pi}}\int_\RR dy e^{-{y^2\over2}}e^{-ty} \ ,
\ee
and then using $e^{-yD}f(x)=f(x-y)$, where $D=d/dx$. This gives the Weierstrass transform of $f$
\be
e^{{D^2\over2}} f(x)={1\over \sqrt{2\pi}}\int_\RR dy e^{-{y^2\over2}} f(x-y) \ .
\label{abbastanzino}\ee
Noticing that this expression is invariant if $f(x-y)$ is replaced by $f(x+y)$, one recognizes it as a particular case of (\ref{weierstrass}).

\noindent
The relation (\ref{daglie}) can be seen as an extension of the relation between the Hermite polynomials and their Weierstrass representation
\be
(-1)^ne^{x^2/2}D^ne^{-x^2/2}=e^{-{D^2/2}} x^n \ .
\label{questaqui}\ee
The left hand side is the standard representation of the so-called probabilistic Hermite polynomial
$He_n$, related to the physicist Hermite polynomial by $H_n(x)=2^{n\over 2} He_n(\sqrt 2 x)$.
Eq.(\ref{questaqui}) is equivalent to
\be
e^{-x^2/2}D^ne^{x^2/2}=e^{{D^2/2}} x^n \ ,
\ee
obtained by replacing $x$ by $ix$ in (\ref{questaqui}).
Given the MacLaurin series
\be
f(x)=\sum_{n=0}^\infty c_n x^n \ ,
\ee
one gets
\be
e^{-x^2/2}f(D)e^{x^2/2}=e^{{D^2/2}} f(x) \ .
\label{onedimension}\ee
Note that this also provides the following suggestive ``perturbative expansion''
\be
e^{-x^2/2}f(D)e^{x^2/2}=e^{{D^2/2}} f(x)=\sum_{n=0}^\infty (-i)^n c_n He_n(ix) \ .
\label{onedimensionpert}\ee
In this respect, we recall that the Hermite polynomials can be written explicitly
\be
He_n(x) = n! \sum_{k=0}^{[\tfrac{n}{2}]} \frac{(-1)^k}{k!(n - 2k)!} \frac{x^{n - 2k}}{2^k} \ .
\ee
Eq.(\ref{onedimensionpert}) can be in fact used in quantum field theory. An obvious reason is that
the dual expression of the generating functional involves $\exp({1\over2} {\delta_{\phi_c}} \Delta{\delta_{\phi_c}})$ acting on a functional
of $\phi_c$. This means that in a perturbative expansion there appear terms such as
\be
\exp\Big({1\over2} {\delta_{\phi_c}} \Delta{\delta_{\phi_c}}\Big) \phi_c^n \ .
\ee
On the other hand,
\be
{\delta_{\phi_c}} \Delta{\delta_{\phi_c}}\phi_c^n(x)=n(n-1)\Delta(0) \phi_c^{n-2}(x) \ ,
\ee
is the functional version of
\be
\Delta(0)\partial_{\phi_c}^2 \phi_c^n=n(n-1)\Delta(0) \phi_c^{n-2} \ ,
\ee
so that, by (\ref{onedimensionpert}),
\be
\exp\Big({1\over2} {\delta_{\phi_c}} \Delta{\delta_{\phi_c}}\Big) \phi_c^n(x)=(-i)^n\Delta^{n\over2}(0) He_n\Big({i\phi_c(x)\over\Delta^{1\over2}(0)}\Big) \ .
\label{eccoquala}\ee

\subsection{Schwinger-Dyson equation in the dual representation}\label{SDSD}

Here we consider the Schwinger-Dyson equation using the dual representation of the generating functional.
In particular, here we consider the case of  the potential ${\lambda^n\over n!}:\phi^n:$. This suggests an operatorial relation that will be derived
in  Sec. \ref{sec:dual}. As we will see, such a relation implies that for arbitrary
normal ordered potentials, the Schwinger-Dyson equation reduces to
\be
\Big[{\delta\over\delta\phi_c(x)}+e^{2U_0[\phi_c]} \int{\delta V\over\delta \phi(x)}\Big(\Delta {\delta\over \delta \phi_c}\Big)e^{\chi \Delta^{-1}\phi_c}|_{\chi=\phi_c}\Big]e^{{1\over2}{\delta\over \delta \phi_c} \Delta{\delta\over\delta \phi_c}}
e^{-\int :V:  (\phi_c)}=0 \ .
\label{SDnewst}\ee
Note that here we used the symbol $:V:$ to denote the normal ordered potential, to distinguish it from the non-normal ordered potential in the square bracket.
The standard form of the Schwinger-Dyson equation
\be
\Big[\Delta^{-1}{\delta\over\delta J}(x)+\int{\delta V\over \delta\phi(x)}\Big({\delta\over\delta J}\Big)-J(x)\Big]W[J]=0 \ ,
\label{lastandard}\ee
expressed in terms of the generating functional $T[\phi_c]$, corresponds to
\be
\Big[{\delta\over\delta\phi_c(x)}+e^{U_0[\phi_c]} \int{\delta V\over\delta \phi(x)}\Big(\Delta {\delta\over \delta \phi_c}\Big)
e^{-U_0[\phi_c]}\Big]e^{{1\over2}{\delta\over \delta \phi_c} \Delta{\delta\over\delta \phi_c}}
e^{-\int V(\phi_c)}=0 \ .
\label{SD}\ee
We now show the connection of the Schwinger-Dyson equation with Hermite polynomials. First note that
Eq.(\ref{daglie}) admits a generalization.
Even if it has been derived by quantum field theoretical methods, it actually depends on quantum objects only through $\Delta(x-y)$.
This indicates that Eq.(\ref{daglie}) is a particular case of a more general relation. Namely, given a function $I$, a functional $F$ and
an even function or distribution $M$, we have the functional generalization of (\ref{onedimension})
\be
\exp\Big(-{1\over2} I M I\Big) F[\delta_I] \exp\Big({1\over2} I M I\Big)= \exp\Big({1\over2} {\delta_I} M^{-1}\delta_I\Big)[FM  I] \ .
\label{generale}\ee
 As we said, later we will prove a more general formula, corresponding to the operatorial extension of (\ref{generale}).

\noindent
The connection with the Hermite polynomials is a consequence of (\ref{eccoquala}) and (\ref{generale}). To see this, note that Eq.(\ref{generale}) implies
\begin{align}
e^{U_0[\phi_c]} {\delta^n\over\delta \phi_c^n(x)} e^{-U_0[\phi_c]} & = \Big[e^{U_0[\phi_c]} \sum_{k=0}^n {n\choose k}
{\delta^{n-k}\over\delta \phi_c^{n-k}(x)}e^{-U_0[\phi_c]}\Big]{\delta^k\over\delta\phi_c^k(x)} \cr
&=\Big[\exp\Big({1\over2} {\delta_{\phi_c}} \Delta{\delta_{\phi_c}}\Big)\sum_{k=0}^n {n\choose k}
(\Delta^{-1} \phi_c)^{n-k}(x)\Big]{\delta^k\over\delta\phi_c^k(x)} \ ,
\label{unexpected}\end{align}
so that, by (\ref{eccoquala}), the Schwinger-Dyson equation for $V={\lambda\over n!}\phi^n$ is
\be
\Big[{\delta\over\delta\phi_c(x)}+\sum_{k=0}^{n-1}{\lambda(-i)^{k}\Delta^{{k}\over2}(0)\over (n-k-1)!k!}  He_{k}
\Big({i\phi_c(x)\over\Delta^{1\over2}(0)}\Big)\Big(\Delta {\delta\over\delta\phi_c}\Big)^{n-k-1}(x)\Big]e^{{1\over2}{\delta\over \delta \phi_c} \Delta{\delta\over\delta \phi_c}}
e^{-\int V(\phi_c)}=0 \ .
\label{tobecompared}\ee

\noindent Let us now consider the potential
\be
: V   (\phi) : = {\lambda\over n!} :\phi^n:   = {\lambda\over n!}\exp\Big(-{1\over2}{\delta\over \delta \phi} \Delta{\delta\over\delta \phi}\Big)\phi^n\ ,
\label{pol}\ee
where we used (\ref{WickTh}) and $TV(\phi)=V(\phi)$.
According to (\ref{SDnewst}), we have
\be
\Big[{\delta\over\delta\phi_c(x)}+{\lambda\over(n-1)!}\sum_{k=0}^{n-1}{n-1\choose k}
\phi_c^k(x)\Big(\Delta {\delta\over\delta\phi_c}\Big)^{n-k-1}(x)\Big]e^{{1\over2}{\delta\over \delta \phi_c} \Delta{\delta\over\delta \phi_c}}
e^{-\int :  V  (\phi_c):}=0 \ ,
\label{nuovo}\ee
that should be compared with (\ref{tobecompared}).
Therefore the terms $e^{U_0[\phi_c]}$ and $e^{-U_0[\phi_c]}$ in the dual representation of the Schwinger-Dyson equation (\ref{SD}), compensate
the contributions coming from the normal ordering regularization of the potential.

\section{Factorization problem}\label{sec-factor}

In this section we show that the dual representation is the natural one to investigate the the following decomposition problem. Namely, given the summation of potentials
\be
V(\phi)=\sum_{k=1}^n V_k(\phi) \ ,
\ee
find how the generating functional associated to $V$ decomposes in terms of the generating functionals associated to the potentials $V_k$'s.
Such an investigation, that may be extended to the case of higher spin fields,  may have several applications, for example in considering
perturbations with respect to a given background. Other applications concern
the analysis of possible symmetries related to a subset of the $V_k$'s.
Here we introduce the method and consider the application to the case of renormalization of scalar theories. Other applications will be investigated in a future
work.

\noindent
Let us consider the simplest case
\be
V(\phi)=V_1(\phi)+V_2(\phi) \ .
\label{unoedue}\ee
In the Schwinger representation we have
\be
W[J]={N\over N_0}\exp\Big(-\int V_1\Big({\delta\over\delta J}\Big)\Big)  \exp\Big(\int V_2\Big({\delta\over\delta J}\Big)\Big) \exp(-Z_0[J]) \ ,
\ee
and there is no an obvious way to understand the structure of the decomposition. In the dual representation of the generating functional, we have
\be
T[\phi_c]={N\over N_0}\exp(-U_0[\phi_c])\exp\Big({1\over2}{\delta\over \delta \phi_c} \Delta{\delta\over\delta \phi_c}\Big) \exp\Big(-\int V_1(\phi_c)\Big) \exp\Big(-\int V_2(\phi_c)\Big) \ ,
\label{totale}\ee
so that, in this case, we can use a key relation satisfied by $\exp\big({1\over2}\delta_{\phi_c}\Delta\delta_{\phi_c}\big)$. Namely, given the functionals $F$ and $G$ we have \cite{FriedBook}
\begin{align}
&\exp\big({{1\over2}  \delta_{\phi_c}\Delta\delta_{\phi_c}}\big)F[\phi_c]G[\phi_c] \cr
&= \exp\big({\delta_{\phi_{c_1}}\Delta\delta_{\phi_{c_2}}}\big)\Big(\exp\big({{1\over2}\delta_{\phi_{c_1}}\Delta\delta_{\phi_{c_1}}}\big)F[\phi_{c_1}]
\exp\big({{1\over2}\delta_{\phi_{c_2}}\Delta\delta_{\phi_{c_2}}}\big) G[\phi_{c_2}]\Big)|_{\phi_{c_1}=\phi_{c_2}=\phi_{c}} \ . \cr
\label{accordings}\end{align}
To derive such a relation, first note the identity
\begin{align}
\exp\big({1\over2} & \delta_{\phi_c}\Delta\delta_{\phi_c}\big)F[\phi_c]G[\phi_c] \cr
& = F[\delta_\mu]G[\delta_\nu]\exp\big({1\over2}\delta_{\phi_c}\Delta\delta_{\phi_c}\big)\exp\Big[\int(\mu+\nu)\phi_c\Big]|_{\mu=\nu=0}\cr
&=F[\delta_\mu]G[\delta_\nu]\exp\Big[{1\over2}\mu\Delta\mu+{1\over2}\nu\Delta\nu+\mu\Delta\nu+\int(\mu+\nu)\phi_c\Big]|_{\mu=\nu=0} \ .
\label{questas}\end{align}
On the other hand, (\ref{accordings}) can be expressed in the form
\begin{align}
 & \exp\big({1\over2} \delta_{\phi_c}\Delta\delta_{\phi_c}\big)F[\phi_c]G[\phi_c] \cr
& = \exp\big(\delta_{\phi_{c_1}}\Delta\delta_{\phi_{c_2}}\big)\Big[F[\delta_\mu]G[\delta_\nu]\exp\big({{1\over2}\mu\Delta\mu+{1\over2}\nu\Delta\nu +\int (\mu\phi_{c_1}+\nu\phi_{c_2})}\big)|_{\mu=\nu=0}\Big]
|_{\phi_{c_1}=\phi_{c_2}=\phi_{c}} \ . \cr
\label{labello}\end{align}
Next observe that, since $\exp\big(\mu\Delta\delta_{\phi_{c}}\big)$ translates a functional of $\phi_{c}$ by $\mu\Delta$, we have
\begin{align}
\exp\big(\delta_{\phi_{c_1}} & \Delta\delta_{\phi_{c_2}}\big)\exp\Big[\int(\mu \phi_{c_1}+\nu\phi_{c_2})\Big] \cr
&=\exp\Big(\int \mu \phi_{c_1}\Big)
\exp\big(\mu\Delta\delta_{\phi_{c2}}\big)\exp\Big(\int \nu \phi_{c2}\Big) \cr
&=\exp\Big[\int (\mu \phi_{c_1}+\nu \phi_{c2}+\mu\Delta
\nu)\Big] \ ,
\end{align}
that applied to (\ref{labello}) reproduces (\ref{questas}). Observe that (\ref{accordings}) implies
the following recursive rule, useful in several computations, e.g. in evaluating the Green's functions
\begin{align}
\exp\Big({1\over2}\delta_{\phi_c} & \Delta\delta_{\phi_c}\Big) \phi_c(x_1)\cdots\phi_c(x_k)|_{\phi_c=0} \cr
&=\sum_{j=2}^k\Delta(x_1-x_j) \exp\Big({1\over2}\delta_{\phi_c} \Delta\delta_{\phi_c}\Big)\phi_c(x_2)\cdots \check\phi_c(x_j)\cdots\phi_c(x_k)|_{\phi_c=0} \ .
\end{align}
Set
\be
\nn F [M  I]  \nn = \exp\Big({1\over2} {\delta_I} M^{-1}\delta_I\Big)F[M  I] \ .
\label{qui}\ee
In this notation, equation (\ref{accordings}) reads
\be
\nn F[\phi_c]G[\phi_c] \nn =\exp(\delta_{\phi_{c_1}}\Delta\delta_{\phi_{c_22}})\nn F[\phi_{c_1}]\nn
\nn G[\phi_{c_2}]\nn|_{\phi_{c_1}=\phi_{c_2}=\phi_{c}} \ .
\label{accordingsb}\ee
Denote by $T_k[\phi_c]$ the generating functional associated to $V_k(\phi)$. For notational reasons we use the same symbol $T_k[\phi_c]$ to denote $T_k[\phi_c]$ divided by the normalization
factor $N_k/N_0$,
$N_k=1/\int D\phi \exp(-S_0-\int V_k(\phi))$, so that
\be
T_k[\phi_c]=\exp(-U_0[\phi_{c}])\exp(\delta_{\phi_{c}}\Delta\delta_{\phi_{c}})\exp\Big(-\int V_k(\phi_c)\Big) \ .
\ee
Similarly, we absorb $N/N_0$ in $T[\phi_c]$.
We then have that the factorization of the generating functional (\ref{totale}) reads
\be
T[\phi_c] =\exp(-U_0[\phi_{c}])\exp(\delta_{\phi_{c_1}}\Delta\delta_{\phi_{c_2}})\nn \exp\Big(-\int V_1(\phi_{c_1})\Big) \nn \nn \exp\Big(-\int V_2(\phi_{c_2})\Big) \nn |_{\phi_{c_1}=\phi_{c_2}=\phi_{c}} \ ,
\ee
that is
\be
T[\phi_c] =  \exp(\delta_{\phi_{c_1}}\Delta\delta_{\phi_{c_2}})\exp\theta(1,2)T_1[\phi_{c_1}]T_2[\phi_{c_2}]|_{\phi_{c_1}=\phi_{c_2}=\phi_{c}} \ ,
\label{quattrodue}\ee
where
\be
\theta(1,2)=-U_0[\phi_{c}]+U_0[\phi_{c_1}]+U_0[\phi_{c_2}] \ .
\label{thetaunodue}\ee
Let us apply Eq.(\ref{quattrodue}) to renormalization. Consider the renormalized action
\be
S_{ren}=\int\Big({1\over 2}\phi\hat\Delta^{-1}\phi+V(\phi)+V_{ct}(\phi)\Big) \ .
\label{renormalized}\ee
where $V_{ct}$ is the counterterm potential and $\hat\Delta$ is the Feynman propagator associated to the full kinetic part of the Lagrangian density
\be
{1\over2}(1+A)\partial_\mu\phi\partial_\mu\phi+{1\over2}(1+B)m^2\phi^2 \ .
\ee
It is understood that now the field $\phi_c$ satisfies the equation of motion
\be
[-(1+A)\partial^2+(1+B)m^2]\phi_c=J \ .
\ee
We denote by $T_{ren}[\phi_c]$ the generating functional associated to $S_{ren}$, by $\hat T[\phi_c]$ the one associated to $S_{ren}-\int V_{ct}$ and by $\hat T_{ct}[\phi_c]$ the one associated to $S_{ren}-\int V$.
By (\ref{quattrodue}), the decomposition of the renormalized generating functional is
\be
T_{ren}[\phi_c] = \exp(\delta_{\phi_{c_1}}\hat \Delta\delta_{\phi_{c_2}})\exp\hat\theta(1,2)\hat T[\phi_{c_1}]\hat T_{ct}[\phi_{c_2}]|_{\phi_{c_1}=\phi_{c_2}=\phi_{c}}  \ ,
\label{op}\ee
where $\hat\theta(1,2)$ is given by (\ref{thetaunodue}) with $\Delta$ replaced by $\hat \Delta$.
Note that Eq.(\ref{op}) is non-perturbative, and can be iterated order by order in the loop expansion.

\noindent We now show that associativity applied to Eq.(\ref{quattrodue}) leads to a relation reminiscent of a cocycle condition. Consider the case of the sum of three potentials
\be
V(\phi)=V_1(\phi)+V_2(\phi)+V_3(\phi) \ ,
\ee
and denote by $T_{jk}[\phi_c]$ the generating functional associated to the potential $V_j(\phi)+V_k(\phi)$, $j,k=1,2,3$. The full generating functional $T[\phi_c]$ can be
derived in the same way of (\ref{quattrodue}). One just imposes the associativity condition by identifying (\ref{quattrodue}), where $V_1$ is replaced by $V_1+V_2$ and $V_2$ by $V_3$,
with the expression obtained replacing  $V_2$ by $V_2+V_3$. This gives
\begin{align}
\exp(\delta_{\phi_{c_{12}}} & \Delta\delta_{\phi_{c_3}})\exp\theta(12,3)T_{12}[\phi_{c_{12}}]T_3[\phi_{c_3}]|_{\phi_{c_{12}}=\phi_{c_3}=\phi_{c}} \cr
& = \exp(\delta_{\phi_{c_{1}}}\Delta\delta_{\phi_{c_{23}}})\exp\theta(1,23)T_{1}[\phi_{c_{1}}]T_{23}[\phi_{c_{23}}]|_{\phi_{c_{1}}=\phi_{c_{23}}=\phi_{c}} \ ,
\end{align}
that, after a change of notation, reads
\be
\exp(\delta_{\phi_{c_{A}}} \Delta\delta_{\phi_{c_B}})\exp\theta(A,B)\Big(T_{12}[\phi_{c_{A}}]T_3[\phi_{c_B}]-T_{1}[\phi_{c_{A}}]T_{23}[\phi_{c_B}]\Big)|_{\phi_{c_{A}}=\phi_{c_B}=\phi_{c}}=0 \ .
\ee

\section{Dual representation and normal ordering}\label{sec:dual}

A feature of the dual representation is that it provides the explicit factorization of the free part $e^{-U_0[\phi_c]}$. The remanent part is the inverse of the
normal ordering operator acting on $e^{-\int V(\phi_c)}$. On the other hand, the Schwinger-Dyson equation in the dual representation led us to consider the relation
(\ref{unexpected}), making clear how $e^{U_0[\phi_c]}$ and its inverse are related to normal ordering. In this respect, note that, how (\ref{unexpected}) shows, Eq.(\ref{generale})
does not hold as an operator relation.
In the following, we will prove that the operatorial version of Eq.(\ref{generale}) is
\be
\exp\Big(-{1\over2} I M I\Big) F[\delta_I] \exp\Big({1\over2} I M I\Big)  =  \exp\Big(-I M I\Big) \nn F [\delta_I]\nn \exp\Big(L M  I\Big)|_{L = I}   \ ,
\label{generalissimabisse}\ee
where $\nn \cdot \nn$ is defined in (\ref{qui}) and $\nn F [\delta_I]\nn$ denotes $\nn F [MI]\nn$ with $MI$ replaced by $\delta_I$.
This relation implies also (\ref{outstanding}).
Eq.(\ref{generale}) is reproduced by acting with (\ref{generalissimabisse}) on a constant  and then noticing that in this case the role of the term $\exp\big(L M  I\big)|_{L = I}$ is to replace
$\nn F  [\delta_I]\nn$ by $\nn F [M  I]\nn$.
Note that replacing $F$ in (\ref{generalissimabisse}) by its normal ordered version
\be
:  F [M  I]:= \exp\big(-{1\over2} {\delta_I} M^{-1}\delta_I\big)F[M  I] \ ,
\ee
yields
\be
\exp\Big(-{1\over2} I M I\Big) :  F [\delta_I] : \exp\Big({1\over2} I M I\Big)  =  \exp\big(-I M I\Big) F[\delta_I]\exp\big(L M  I\Big)|_{L = I}   \ .
\label{generalissima}\ee
\noindent It is immediate to check that (\ref{generalissima}) applied to (\ref{SD}), with $V$ replaced by its normal ordered version, leads to (\ref{SDnewst}).

\noindent In this section, we first prove (\ref{generalissima}), that implies also (\ref{generalissimabisse}), then we show that this is the functional analog of a relation satisfied
by the Hermite polynomials.

\subsection{Proof of (\ref{generalissima})}

To prove (\ref{generalissima}) we first express
$:  F  [M  I]:$ in terms of $e^{-{1\over2} {\delta_I} M^{-1}\delta_I}$ acting on the Laplace transform of $F[M  I]$
\be
:  F  [M  I] : = e^{-{1\over2} {\delta_I} M^{-1}\delta_I}\int DJ e^{IMJ} \hat F[M  J]=\int DJ e^{-{1\over2} J M J+IMJ} \hat F[M  J] \ ,
\ee
so that
\be
:  F  [\delta_I] : = \int DJ e^{-{1\over2} J M J+J\delta_I} \hat F[M  J] \ .
\ee
Then, we act with (\ref{generalissima}) on a functional $G[MI]$. Since
\begin{align}
:  F [\delta_I] : e^{{1\over2} I M I}G[MI]&=\int DJ e^{-{1\over2} J M J+J\delta_I} \hat F[M  J] e^{{1\over2} I M I} G[MI] \cr
&=\int DJ e^{-{1\over2} J M J} \hat F[M  J] e^{{1\over2} (I+J) M (I+J)} G[M(I+J)] \ ,
\end{align}
it follows that
\begin{align}
e^{-{1\over2} I M I} :  F [\delta_I] : e^{{1\over2} I M I}G[MI]&=\int DJ e^{IMJ}\hat F[M  J] G[M(I+J)] \cr
&= e^{-I M I}\int DJ \hat F[M  J]e^{LM(I+J)}|_{L=I} G[M(I+J)] \cr
&= e^{-I M I}\int DJ e^{J\delta_I} \hat F[M  J] e^{LMI}|_{L=I} G[MI] \cr
&= e^{-I M I}F[\delta_I]e^{LMI}|_{L=I} G[MI] \ ,
\end{align}
which is (\ref{generalissima}) acting on $G[MI]$.

\subsection{Operatorial extension of the Weierstrass representation of the Hermite polynomials}

Eq.(\ref{generalissima}) is the functional generalization of the identity
\be
e^{-x^2/2}\Big(e^{-{D^2/ 2}}f(x)\Big)(D) e^{x^2/2}=e^{-x^2}f(D) e^{yx}|_{y=x} \ ,
\label{unadim}\ee
$D=d/dx$. Replacing $f(x)$ by $e^{{D^2/ 2}}f(x)$, Eq.(\ref{unadim}) reads
\be
e^{-x^2/2}f(D) e^{x^2/2}=e^{-x^2}\Big(e^{{D^2/ 2}}f(x)\Big)(D) e^{yx}|_{y=x} \ .
\label{unadimW}\ee
Acting with (\ref{unadimW}) on a constant yields
\be
e^{-x^2/2}f(D) e^{x^2/2}=e^{{D^2/ 2}}f(x) \ .
\label{unadimWsss}\ee
In the case $f(x)=x^n$, Eq.(\ref{unadimW}) provides the operatorial extension of the Weierstrass representation of the Hermite polynomials.

\noindent
The proof of (\ref{unadim}) can be done by expressing $f$ in terms of its Laplace transform as done for
(\ref{generalissima}). Nevertheless, it is of interest to stress its connection with the Hermite polynomials. We then consider the case
$f(x)=x^n$, with $n$ a non-negative integer.
We do this using the induction method. Since (\ref{unadim}) holds for $f(x)=x$ and
\be
e^{-{D^2/2}}x^k=H_k(x) \ ,
\ee
we should prove the operatorial relation
\be
e^{-{x^2/2}}H_{k}(D)e^{x^2/2}=e^{-x^2}D^{k} e^{yx}|_{y=x} \ ,
\label{npiuuno}\ee
for $k=n+1$, assuming that it holds for $k=n$.
On the other hand,
\begin{align}
e^{-x^2}D^{n+1} e^{yx}|_{y=x}&=e^{-x^2}yD^{n} e^{yx}|_{y=x}+e^{-x^2}D^{n} e^{yx}|_{y=x}D \cr
&= e^{-x^2}x D^{n} e^{yx}|_{y=x}+e^{-x^2}D^{n} e^{yx}|_{y=x}D \ .
\end{align}
Therefore, since
\be
H_{n+1}(x)=xH_n(x)-nH_{n-1}(x)\ ,
\ee
it remains to prove that
\be
e^{-{x^2/2}}(H_{n}(D)D-nH_{n-1}(D))e^{x^2/2}=e^{-x^2}x D^{n} e^{yx}|_{y=x}+e^{-x^2}D^{n} e^{yx}|_{y=x}D \ .
\label{edfg}\ee
To this end, note that
\begin{align}
e^{-{x^2/2}}H_{n}(D)De^{x^2/2} & = e^{-{x^2/2}}H_{n}(D)xe^{x^2/2}+e^{-{x^2/2}}H_{n}(D)e^{x^2/2}D \cr
& = e^{-{x^2/2}}(xH_{n}(D)+nH_{n-1}(D))e^{x^2/2}+e^{-{x^2/2}}H_{n}(D)e^{x^2/2}D \ ,
\label{hgb}\end{align}
where in the last equality we used
\be
[H_n(D),x]=nH_{n-1}(D) \ .
\ee
By (\ref{edfg}) and (\ref{hgb}) we get
\be
e^{-{x^2/2}}xH_{n}(D)e^{x^2/2}+e^{-{x^2/2}}H_{n}(D)e^{x^2/2}D=e^{-x^2}x D^{n} e^{yx}|_{y=x}+e^{-x^2}D^{n} e^{yx}|_{y=x}D \ ,
\ee
which is the assumption.

\section{$T[\phi_c]$ and normal ordered potentials}\label{sec-five}

Here we first investigate the structure of the generating functional $U[\phi_c]=-\ln T[\phi_c]$ in the case of normal ordered potentials.
This is done by specializing the construction in \cite{FriedBook}. In particular, the action of the inverse of the Wick operator on each potential, contributing to the expansion,
precisely cancels the normal ordering, so that
leading to a considerable simplification of the full series of $U[\phi_c]=-\ln T[\phi_c]$.
Comparing the result with the explicit expression of $W[J]$ in the case of the potential ${\lambda\over n!}:\phi^n:$, recently derived in \cite{Matone:2015nxa},
yields an identity that implies an infinite set of relations.
We then derive such relations using the Fa\`a di Bruno formula, concerning the chain rules for higher order derivatives. This provides, for all $n$, the
Feynman combinatorics and
the explicit form of the full loop expansion of $U[\phi_c]$. We conclude the section by showing a duality between the field $\phi$ and the external source $\phi_c$. Promoting
$\phi_c$ to a dynamical field leads
to a scalar theory with potential $V(\phi+\phi_c)$, which is described by the theory of a single field $\phi_c$ with the generating functional $U[\phi_c]$ playing the role
of potential.

\subsection{The general case}

In the following we consider normal ordered potentials that, for notational reasons, we denote by $:V:$.
Let us set
\be
\D = {1\over2}{\delta\over \delta \phi_c} \Delta{\delta\over\delta \phi_c} \ ,
\ee
so that
\be
T[\phi_c]= {N\over N_0} \exp(-U_0 [\phi_c]) \exp(D) \exp\Big(-\int : V(\phi_c) :\Big) \ .
\label{labm}\ee
We also define
\be
\D_j={1\over2}{\delta\over \delta \phi_{c_j}} \Delta{\delta\over\delta \phi_{c_j}} \ , \qquad \D_{jk}={\delta\over \delta \phi_{c_j}} \Delta{\delta\over\delta \phi_{c_k}}  \ ,
\ee
$j,k\in \NN_+$.
Let us rewrite (\ref{accordings}) in this notation
\be
e^\D F[\phi_c] G[\phi_c]=e^{\D_{12}}\Big(e^{\D_1} F[\phi_{c_1}]e^{\D_2}G[\phi_{c_2}]\Big)|_{\phi_{c_1}=\phi_{c_2}=\phi_{c}} \ ,
\ee
We set $U[\phi_c]=Z[J]$, and write
\be
T[\phi_c]=\exp(-U[\phi_c])={N\over N_0}\exp\Big(-U_0[\phi_c]+\sum_{k=1}^\infty {Q_k[\phi_c]\over k!}\Big) \ .
\label{labn}\ee
Rescaling the potential by a constant $\mu$, and then expanding $\exp(-\smallint : V :  )$ in (\ref{labm}), gives by (\ref{labn})
\be
\exp(\D)\exp\Big(-\mu \int : V(\phi_c) :\Big)=\exp\Big(\sum_{k=1}^\infty {\mu^k\over k!}Q_k[\phi_c]\Big) \ ,
\ee
so that
\be
Q_k[\phi_c]= \partial_\mu^k \ln \Big[\exp(\D)\exp\Big(-\mu \int :V(\phi_c):\Big)\Big]|_{\mu=0} \ .
\ee
There is a considerable simplification since the normal ordering cancels all the $e^{\D_k}$s. The first two cases are
\be
Q_1= - e^\D\int :  V : = - \int V   \ ,
\ee
and
\begin{align}
Q_2 &=\Big[e^{\D_{12}}-1\Big]\Big[\Big(e^{\D_1} \int :  V(\phi_{c_1}):  \Big)\Big(e^{\D_2} \int :  V(\phi_{c_2}):  \Big)\Big]|_{\phi_{c_1}=\phi_{c_2}=\phi_{c}} \cr
&=\Big[e^{\D_{12}}-1\Big]\Big[\int V(\phi_{c_1}) \int V(\phi_{c_2})\Big]|_{\phi_{c_1}=\phi_{c_2}=\phi_{c}} \ .
\end{align}
Note that, as seen by expanding $e^{\D_{12}}$, the effect of the $-1$ in $e^{\D_{12}}-1$ is to eliminate Feynman diagrams which are not connected by at least one propagator.

\noindent
As in the case of $Q_1$ and $Q_2$, all the operators $e^{\D_k}$, $k=1,\ldots,n$,
disappear from the expression of the $Q_n$s and their expression simplify to
\be
Q_n[\phi_c]=(-1)^n\prod_{j>k}^n e^{\D_{jk}}\prod_{i=1}^n \int V(\phi_{ci})|_{c, \phi_{c_1},\ldots,\phi_{c_n}=\phi_c} \ ,
\label{leqenne}\ee
where the subscript $c$ means that terms non-connected by at least one Feynman propagator should be discarded.

\noindent It then follows by (\ref{labn}) and (\ref{leqenne}) that the generating functionals of connected Green's functions associated to a normal ordered potential is
\be
U[\phi_c]=\ln{N_0\over N}+U_0[\phi_c]+\sum_{p=1}^\infty {(-1)^{p+1}\over p!}\prod_{j>k}^p e^{\D_{jk}}\prod_{i=1}^p \int V(\phi_{c_i})|_{c, \phi_{c_1},\ldots,\phi_{c_p}=\phi_c} \ .
\label{UUUUUU}\ee

\subsection{A combinatorial identity for ${\lambda\over n!}:  \phi^n:  $}

In \cite{Matone:2015nxa} it has been introduced a method to derive the generating functional
of scalar potentials starting from the one associated to the exponential potential $\mu^D\exp(\alpha\phi)$, seen as a master potential. In particular, it has been shown how
this leads to derive the generating functional for normal ordered potentials, absorbing at once all the $\Delta(0)$ terms, by using
\be
:  \exp(\alpha\phi) :   = \exp\Big(-{\alpha^2\over2}\Delta(0)\Big) \exp(\alpha\phi) \ .
\ee
It turns out that in the case of $:  V(\phi):  ={\lambda\over n!}:  \phi^n:  $, the generating functional $T^{(n)}[\phi_c]$, and therefore all the Green's functions, can be easily expressed in the explicit form.
Let us introduce the symbol
\be
\sum_{p,q}^{n,k}\equiv \sum_{p=0}^{[{kn\over2}]} \sum_{\sum_{i=1}^k q_i=kn-2p}\sum_{p_1=\ldots=p_k=n}  \ ,
\label{symbol}\ee
where $[a]$ denotes the integer part of $a$ and $0\leq q_i\leq kn-2p$.
We have  \cite{Matone:2015nxa}
\be
T^{(n)}[\phi_c] =
{N\over N_0}e^{-U_0[\phi_c]}\sum_{k=0}^\infty{{(-\lambda)^k\over k!}}\sum_{p,q}^{n,k}
[k|m,q]\prod_{i=1}^k\int d^Dz_i\phi_c^{q_i}(z_i)\prod_{l>j}^k\Delta(z_j-z_l)^{m_{jl}} \ ,
\label{bois}\ee
where
\be
[k|m,q]={1\over \prod_{i=1}^k q_{i}! \prod_{l>j}^k m_{lj}!} \ ,
\ee
with the $m_{lj}$'s taking all the values satisfying the conditions  $0\leq m_{lj}\leq p$ and $\sum_{l>j=1}^km_{lj}=p$. Furthermore,
\be
p_l=\sum_{i=1}^{l-1}m_{il}+\sum_{j=l+1}^k m_{lj}+q_l \ ,
\label{pelleforj}\ee
$l=1,\ldots,n$.

\noindent
By (\ref{labn}), (\ref{UUUUUU}) and (\ref{bois}), we get the non-trivial identity
\begin{align}
&\exp\Big[\sum_{p=1}^\infty {(-\lambda)^{p}\over p!(n!)^p}\prod_{j>k}^p e^{\D_{jk}}\prod_{i=1}^p \int d^Dz_l \phi_{ci}^n(z_l)|_{c, \phi_{c_1},\ldots,\phi_{c_p}=\phi_c}\Big] \cr
&=\sum_{k=0}^\infty{{(-\lambda)^k\over k!}}\sum_{p,q}^{n,k}
[k|m,q]\prod_{i=1}^k\int d^Dz_i\phi_c^{q_i}(z_i)\prod_{l>j}^k\Delta(z_j-z_l)^{m_{jl}} \ .
\label{boisbis}\end{align}
Comparing the coefficients of all powers of $\lambda$ we will derive the explicit expressions of the action of the product of the $e^{\D_{jk}}$s operators.

\noindent
 $U[\phi_c]$ generates the connected $N$-point functions,
without the free external legs, of $\phi$.  Therefore,
\be
-{\delta^N U[\phi_c]\over\delta \phi_c(x_1)\ldots\delta \phi_c(x_N)}|_{\phi_c=0}=\int \big(\prod_{k=1}^N d^Dy_k\Delta^{-1}(x_k-y_k)\big)\langle 0| T \phi(y_1)\ldots \phi(y_N)|0\rangle_c \ .
\ee
In the case of $n$ odd one has, perturbatively, $\langle\phi(x)\rangle\neq0$. We recall that  for $N\geq2$ such correlators coincide with the ones of
$\eta(x)=\phi(x)-\langle\phi(x)\rangle$ (see, for example, \cite{Matone:2015nxa})
\be
\langle 0| T \phi_(y_1)\ldots \phi(y_N)|0\rangle_c=\langle 0| T \eta(y_1)\ldots\eta(y_N)|0\rangle_c \ .
\ee

\subsection{Feynman combinatorics and Fa\`a di Bruno formula}

We now derive the infinitely many relations implied by (\ref{boisbis}), giving the
explicit form of the action of the operators $e^{\D_{jk}}$. As we will see, this also implies the
following explicit expression for the generating functional $U[\phi_c]$,
in the case of the potential $V={\lambda\over n!}:\phi^n:$, for all $n$
\be
U[\phi_c] = \ln {N_0\over N}+U_0[\phi_c]+\sum_{p=1}^\infty (-\lambda)^p\sum_{l=1}^p (-1)^l\sum_{j_1+\cdots+j_l=p} {h_{n,j_1}\over j_1!}\cdots {h_{n,j_l}\over j_l!}\ ,
\label{laucompatta}\ee
where $j_1,\ldots,j_l\geq1$, and
\be
h_{n,k}=\sum_{p,q}^{n,k}
[k|m,q]\prod_{i=1}^k\int d^Dz_i\phi_c^{q_i}(z_i)\prod_{l>j}^k\Delta(z_j-z_l)^{m_{jl}} \ .
\label{hnk}\ee
Note that in this notation
\be
T^{(n)}[\phi_c]={N\over N_0}\exp(-U_0[\phi_c])\Big(1+\sum_{k=1}^\infty {(-\lambda)^k\over k!}h_{n,k}\Big) \ .
\ee
The derivation follows by the Fa\`a di Bruno formula for the chain rules in the case of higher derivatives. There are several versions
of such a formula. The original one is
\be
{d^m\over dx^m}f(g(x))=\sum_{k_1,\ldots,k_m}{m!\over k_1! \cdots k_m!} f^{(k)}(g(x))\Big({g^{(1)}(x)\over 1!}\Big)^{k_1}
\cdots \Big({g^{(m)}(x)\over m!}\Big)^{k_m} \ ,
\ee
where the sum is over all the nonnegative integer solutions of the Diophantine equation
\be
\sum_{j=1}^mjk_j=m \ ,
\ee
and $k:=\sum_{j=1}^m k_j$.
Here we use the equivalent expression
\be
{d^m\over dx^m}f(g(x))=m!\sum_{l=1}^m {f^{(l)}(g(x))\over l!}\sum_{j_1+\cdots+j_l=m}{g^{(j_1)}(x)\over j_1!}\cdots {g^{(j_l)}(x)\over j_l!} \ ,
\label{questaqua}\ee
with $j_1,\ldots, j_k\geq 1$.
Eq.(\ref{boisbis}) implies that for all positive integers $p$
\be
 {1\over (-n!)^p}\prod_{j>k}^p e^{\D_{jk}}\prod_{i=1}^p \int d^Dz_l \phi_{ci}^n(z_l)|_{c, \phi_{c_1},\ldots,\phi_{c_p}=\phi_c}={d^p\over d\lambda^p}\ln \Big(1+\sum_{k=1}^\infty {(-\lambda)^k\over k!}h_{n,k}\Big)|_{\lambda=0} \ .
\ee
Applying the Fa\`a di Bruno formula (\ref{questaqua}), yields
\be
 \prod_{j>k}^p e^{\D_{jk}}\prod_{i=1}^p \int d^Dz_l \phi_{ci}^n(z_l)|_{c, \phi_{c_1},\ldots,\phi_{c_p}=\phi_c}= (n!)^p p!\sum_{l=1}^p (-1)^{l+1}\sum_{j_1+\cdots+j_l=p} {h_{n,j_1}\over j_1!}\cdots {h_{n,j_l}\over j_l!}\ ,
\label{leh}\ee
and Eq.(\ref{laucompatta}) immediately follows by (\ref{UUUUUU}).

\noindent
Let us compute, for each $l$, the total exponent $E_l$ of $\phi_c$ in the term
\be
\sum_{j_1+\cdots+j_l=p} {h_{n,j_1}\over j_1!}\cdots {h_{n,j_l}\over j_l!} \ ,
\label{528}\ee
of Eq.(\ref{laucompatta}). By (\ref{symbol}) and (\ref{hnk}) it follows that the total exponent of $\phi_c$ in $h_{n,k}$ is
\be
\sum_{i=1}^k q_i=kn-2s \ ,
\ee
with $s$ ranging between $0$ and $[kn/2]$. It follows that
\be
E_l=\sum_{i=1}^l(j_in-2s_i) \ ,
\ee
with $s_i$ ranging between $0$ and $[j_i/2]$. On the other hand, since $\sum_{i=1}^l j_i=p$, we get
\be
E_l=np-2\sum_{i=1}^ls_i \ .
\ee
It follows that the only terms in (\ref{528}) contributing to the connected $N$-point function at $p$-loops, $p\geq 1$, are the ones satisfying
\be
\sum_{i=1}^ls_i={1\over2}(np-N) \ ,
\label{eccol}\ee
which implies
\be
p\geq {N\over n} \ .
\ee
Note that Eq.(\ref{eccol}) also reproduces the well known fact that, in the case of $n$ even, there are contributions at any loop to the $N$-point function with $N$ even.
Furthermore, for $n$ odd, non-vanishing contributions may arise only if $N$ and $p$ have the same parity. Further constraints may be derived by considering the
structure of (\ref{laucompatta}). For example, one may check that there are no contributions of order $p=1$ to the 2-point function, unless in the trivial case $n=2$.
This is just a property of the normal ordered potential ${\lambda\over n!}:\phi^n:$.

\subsection{$\phi$-$\phi_c$ duality and $U[\phi_c]$ as an effective potential}\label{dualphi}

We now show a duality between the field $\phi$ and the source $\phi_c$. This naturally leads to promote $\phi_c$ to a dynamical field, with potential $U[\phi_c]$,
that is with the action
\be
S_D[\phi_c]={1\over2}\phi_c\Delta^{-1}\phi_c+U[\phi_c] \ .
\label{playing}\ee

\noindent Consider the functional
\be
H[\phi,\phi_c]=\exp\Big(-{1\over2}\phi\Delta^{-1}\phi+{1\over2}\phi_c\Delta^{-1}\phi_c-\int V(\phi+\phi_c)\Big) \ .
\label{showsssnew}\ee
According to (\ref{showsss}) we have
\be
T[\phi_c]=N\int D\phi H[\phi,\phi_c] \ .
\label{rto}\ee
Define
\be
O[\phi]=N_c \int D\phi_c H[\phi,\phi_c] \ ,
\label{OOO}\ee
with
\be
N_c^{-1}=\int D\phi_c \exp\Big({1\over2}\phi_c\Delta^{-1}\phi_c-\int V(\phi_c)\Big) \ .
\ee
In spite of the opposite sign of the kinetic term for $\phi_c$, it is worth noticing the exchange
of roles between the external source $\phi_c$ and the scalar field $\phi$ in (\ref{rto}) and (\ref{OOO}).

\noindent
Let us consider a slightly modified version of (\ref{showsssnew}). Namely, define
\be
K[\phi,\phi_c]=  \exp\big(-\phi_c\Delta^{-1}\phi_c\big) H[\phi,\phi_c]=\exp\Big(-{1\over2}\phi\Delta^{-1}\phi-{1\over2}\phi_c\Delta^{-1}\phi_c-\int V(\phi+\phi_c)\Big) \ .
\label{showsssnewdue}\ee
Note that by (\ref{labn}), (\ref{playing}) and (\ref{showsssnewdue}), we have
\be
\exp(-S_D[\phi_c])= N \int D\phi  K[\phi,\phi_c] = \exp\big(-\phi_c\Delta^{-1}\phi_c\big) T[\phi_c] \ ,
\ee
and
\be
S_D[\phi_c]=\ln{N_0\over N}+{1\over2}\phi_c\Delta^{-1}\phi_c-\sum_{k=1}^\infty {Q_k[\phi_c]\over k!} \ .
\label{essedi}\ee
Consider the dual generating functional
\begin{align}
T_D[\varphi_c]&=N_D N \int D\phi_c\int D\phi  K[\phi,\phi_c]\exp (\varphi_c \Delta^{-1}\phi_c) \cr
&=N_D  \int D\phi_c\exp(-S_D[\phi_c]+\varphi_c \Delta^{-1}\phi_c) \ ,
\label{varfi}\end{align}
where $N_D=1/\int D\phi_c \exp(-S_D[\phi_c])$. $T_D[\varphi_c]$ is the generating functional for two scalar fields, $\phi$ and $\phi_c$, symmetrically coupled by the potential $V(\phi+\phi_c)$,
with one of the two external currents set to zero. That is
\be
T_D[\varphi_c]=W[0,\varphi_c\Delta^{-1}] \ ,
\label{NY}\ee
where
\be
W[I,K]=N_DN\int D\phi_c D\phi\exp\Big[-{1\over2}\phi\Delta^{-1}\phi-{1\over2}\phi_c\Delta^{-1}\phi_c -\int (V(\phi+\phi_c)-I\phi-K\phi_c)\Big] \ ,
\label{letusshow}\ee
which is symmetric, that is $W[I,K]=W[K,I]$.

\noindent
Let us consider, for arbitrary $I$ and $K=\varphi_c\Delta^{-1}$, the path integral over $\phi$ and $\phi_c$ in (\ref{letusshow}). This also
gives the expression of the path integration over $\phi_c$ in (\ref{varfi}). To this end,
 we set
\be
\phi=\phi'+\Delta I \ , \qquad \phi_c=\phi_c'+\Delta K \ ,
\ee
in (\ref{letusshow}), and then drop the prime from $\phi'$ and $\phi_c'$, to get
\begin{align}
W[I,K]&=N_DN\exp({1\over2}I\Delta I+{1\over2}K\Delta K)\cr
&\int D\phi_c D\phi\exp\Big[-{1\over2}\phi\Delta^{-1}\phi-{1\over2}\phi_c\Delta^{-1}\phi_c -\int V(\phi+\phi_c+\Delta(I+K))\Big] \ . \cr
\end{align}
Next, defining
\be
\rho={\phi+\phi_c\over \sqrt2} \ , \qquad \sigma={\phi-\phi_c\over \sqrt2} \ ,
\label{rhoesigma}\ee
we get
\be
W[I,K]={N_DN\over N_0}\exp({1\over2}I\Delta I+{1\over2}K\Delta K)\int D\rho \exp\Big[-{1\over2}\rho\Delta^{-1}\rho -\int V(\sqrt2\rho+\Delta(I+K))\Big] \ .
\ee
We then set
\be
\rho_c= {\Delta(I+K)\over\sqrt2} \ , \qquad \sigma_c= {\Delta(I-K)\over\sqrt2} \ ,
\ee
to get $T_D[\rho_c,\sigma_c]=W[I,K]$, with
\be
T_D[\rho_c,\sigma_c]= {N_DN\over N_0^2}\exp({1\over2}\sigma_c\Delta^{-1}\sigma_c+{1\over2}\rho_c\Delta^{-1} \rho_c)\exp\Big({1\over2}\delta_{\rho_c}\Delta\delta_{\rho_c}\Big)\exp\Big(-\int V(\sqrt2\rho_c)\Big) \ .
\label{nidentity}\ee
According to (\ref{labn}) we have
\be
T_D[\rho_c,\sigma_c]= {N_DN\over N_0^2}\exp({1\over2}\sigma_c\Delta^{-1}\sigma_c+{1\over2}\rho_c\Delta^{-1} \rho_c)\exp\Big(\sum_{k=1}^\infty {P_k[\rho_c]\over k!}\Big) \ ,
\ee
where
\be
P_k[\rho_c]=\partial_\mu^k\ln\Big[ \exp\Big({1\over2}\delta_{\rho_c}\Delta\delta_{\rho_c}\Big)\Big(-\mu\int V(\sqrt2\rho_c)\Big)\Big]|_{\mu=0} \ .
\ee
By construction $T_D[\varphi_c]$ corresponds to $T_D[\rho_c,\sigma_c]$ evaluated at $I=0$
\be
T_D[\varphi_c]=T_D[{\varphi_c\over\sqrt2},-{\varphi_c\over\sqrt2}] \ ,
\ee
that, by (\ref{varfi}) and (\ref{nidentity}), corresponds to the identity
\begin{align}
\int D\phi_c & \exp\Big(-{1\over2}\phi_c\Delta^{-1}\phi_c+\sum_{k=1}^\infty {Q_k[\phi_c]\over k!}+\varphi_c\Delta^{-1}\phi_c\Big)\cr
& ={1\over N_0}\exp({1\over2}\varphi_c\Delta^{-1} \varphi_c)\exp(\delta_{\varphi_c}\Delta\delta_{\varphi_c})\exp\Big(-\int V(\varphi_c)\Big) \ ,
\label{eccof}\end{align}
which has the same form of the dual generating functional associated to the potential $V(\phi_c)$ except for the rescaling by a factor 2, of $\delta_{\varphi_c}\Delta\delta_{\varphi_c}$.
This is equivalent to rescale the kinetic term by a factor $1/2$. To see this, note that (\ref{eccof}) can be directly obtained by replacing $\rho$ and $\sigma$ in (\ref{rhoesigma}), by
their rescaled version
\be
\tilde \rho=\phi+\phi_c \ , \qquad \tilde\sigma=\phi-\phi_c \ .
\ee
This leads to
\be
T_D[\varphi_c]= {N_DN\over N_0 N_0'} \exp({1\over2}\varphi_c\Delta^{-1}\varphi_c)\int D\tilde\rho \exp\Big(-{1\over4}\tilde\rho\Delta^{-1}\tilde\rho-\int V(\tilde\rho+\varphi_c)\Big) \ ,
\ee
with $N_0'=1/\int D\tilde\sigma \exp(-{1\over4}\tilde\sigma\Delta^{-1}\tilde\sigma)$.

\noindent Instead of varying the coefficient of the kinetic term, one may rescale $\tilde \rho$ and $\varphi_c$  to get the standard
normalization of the kinetic term, so that replacing $V(\tilde\rho+\varphi_c)$ by
$V(\sqrt2(\tilde\rho+\varphi_c))$. Iterating $n$-times the procedure mapping $T[\phi]$ to $T_D[\varphi_c]$, leads to a rescaling of a factor $2^{n/2}$ of the scalar field. Such an iteration may
lead to an interpretation of the rescaling of the scalar field in the renormalization procedure.

\section{Generating functional and covariant derivatives}\label{covariant}

In this section we derive a new representation of the generating functional, expressed in terms of covariant derivatives acting on 1.
The key observation is that some of the relations we derived in the previous sections, can be expressed in terms
of covariant derivatives. This leads to some new results in the path-integral approach to quantum field theory.

\noindent
The starting point is to note that, given a functional $F$, one has the operator identity
\be
\exp\Big(-{1\over2} I M I\Big) F[\delta_I] \exp\Big({1\over2} I M I\Big)  =  F[{\cal D}_{MI}]   \ ,
\label{generalissimabisseAAA}\ee
where ${\cal D}_{MI}(x)$ denotes the ``covariant derivative''
\be
{\cal D}_{MI}(x)={\delta\over\delta I(x)}+MI(x) \ .
\ee
Eq.(\ref{generalissimabisseAAA}) is the functional generalization of the
operator relation
\be
e^{-x^2/2}f(D) e^{x^2/2}=f(D+x) \ .
\label{unadimWZX}\ee
By (\ref{generale}), which is not an operator identity, and (\ref{generalissimabisseAAA}), it follows that
\be
\exp\Big({1\over2} \delta_I M^{-1} \delta_I\Big) F[MI]  =  F[{\cal D}_{MI}] \cdot 1   \ ,
\label{babablaxy}\ee
which is the functional extension of
\be
e^{D^2/2} f(x)=f(D+x)\cdot 1 \ .
\ee
Furthermore, by (\ref{generalissimabisse}) and (\ref{generalissimabisseAAA}) we get the operator relation
\be
\exp\Big(-I M I\Big) \nn F [\delta_I]\nn \exp\Big(L M  I\Big)|_{L = I}= F[{\cal D}_{MI}] \ .
\ee
By (\ref{generalissimabisseAAA}) we have
\be
\exp(Z_0[J]) \exp\Big(-\int V\Big({\delta\over\delta J}\Big)\Big) \exp(-Z_0[J]) =
\exp\Big(-\int V({\cal D}_{\phi_c}^-)\Big) \ ,
\ee
where
\be
{\cal D}_{\phi}^{\pm}(x)=\mp\Delta{\delta\over\delta\phi}(x)+\phi(x) \ .
\ee
Such operators satisfy the commutation relations
\be
[{\cal D}_{\phi}^{-}(x), {\cal D}_{\phi}^{+}(y)]=2\Delta(x-y) \ ,
\ee
and
\be
[{\cal D}_{\phi}^{-}(x), {\cal D}_{\phi}^{-}(y)]=[{\cal D}_{\phi}^{+}(x), {\cal D}_{\phi}^{+}(y)]=0 \ .
\ee
It follows that $T[\phi_c]$ can be expressed in terms of covariant derivatives acting on 1
\be
T[\phi_c] = {N\over N_0} \exp(-U_0[\phi_c])\exp\Big(-\int V({\cal D}_{\phi_c}^-)\Big) \cdot 1 \ .
\label{nuova}\ee
By (\ref{babablaxy}) or, equivalently, by (\ref{quindici}) and (\ref{nuova}), we have
\be
\exp\Big(\pm{1\over2}{\delta\over\delta\phi_c}\Delta{\delta\over\delta\phi_c}\Big)\exp\Big(-\int V(\phi_c)\Big) = \exp\Big(-\int V({\cal D}_{\phi_c}^\mp)\Big) \cdot 1 \ ,
\label{basicid}\ee
where the action of the Wick operator $\exp(-{1\over2}\delta_{\phi_c}\Delta\delta_{\phi_c})$ has been obtained by changing sign to $\Delta$.
Since
\be
{\delta\over \delta J(x)} \exp(-U_0[\phi_c]) = \exp(-U_0[\phi_c]) {\cal D}_{\phi_c}^-(x) \ ,
\ee
it follows that even the Green's functions can be expressed in terms of the covariant derivatives
\begin{align}
{\delta^N W[J]\over \delta J(x_1) \ldots   \delta J(x_N)}&=\exp(-U_0[\phi_c]){\cal D}_{\phi_c}^-(x_1)\ldots {\cal D}_{\phi_c}^-(x_N)\exp\Big(-\int V({\cal D}_{\phi_c}^-)\Big)\cdot 1 \cr
&= \exp(-U_0[\phi_c])\exp\Big(-\int V({\cal D}_{\phi_c}^-)\Big) {\cal D}_{\phi_c}^-(x_1)\ldots {\cal D}_{\phi_c}^-(x_N) \cdot 1 \ .
\label{firstrow}\end{align}
By (\ref{SD}) and (\ref{basicid}) it follows that the Schwinger-Dyson equation reduces to the identity
\be
\Big({\delta\over\delta\phi_c(x)}+\int{\delta V({\cal D}_{\phi_c}^-)\over\delta \phi(x)}\Big)\exp\Big(-\int V({\cal D}_{\phi_c}^-)\Big) \cdot 1 = 0 \ .
\label{SDpnnews}\ee
The above representation of the generating functional simplifies the explicit calculations. For example, in the case of $V={\lambda\over4!}\phi^4$, one immediately gets
\begin{align}
T[\phi_c]&={N\over N_0}\exp(-U_0[\phi_c])\Big(1-{\lambda\over4!}\int d^Dx {{\cal D}^-_{\phi_c}}^4 (x)+\ldots\Big)\cdot 1 \cr
& = {N\over N_0}\exp(-U_0[\phi_c])\Big[1-{\lambda\over4!}\int d^Dx(\phi_c^4(x)+6\phi_c^2(x)\Delta(0)+3\Delta^2(0))+\ldots\Big] \ ,
\end{align}
Using  ${\cal D}^-_{\phi}(x)\cdot 1=\phi(x)$ and
\be
[{\cal D}^-_{\phi}(x),\phi(y)]=\Delta(x-y) \ ,
\ee
one gets
\begin{align}
\prod_{k=1}^2 &{\cal D}^-_{\phi} (x_k)\cdot 1  =\prod_{k=1}^2 \phi(x_k)+\Delta(x_1-x_2) \ , \cr
\prod_{k=1}^3 &{\cal D}^-_{\phi} (x_k)\cdot 1 = \prod_{k=1}^3\phi(x_1)+\Delta(x_1-x_2)\phi(x_3)+\Delta(x_1-x_3)\phi(x_2)+\Delta(x_2-x_3)\phi(x_1)\ , \cr
\prod_{k=1}^4  &{\cal D}^-_{\phi}(x_k)\cdot 1  = \prod_{k=1}^4\phi(x_k)+\Delta(x_1-x_2)\phi(x_3)\phi(x_4)
+\Delta(x_1-x_3)\phi(x_2)\phi(x_4) \cr
& +\Delta(x_1-x_4)\phi(x_2)\phi(x_3)+\Delta(x_2-x_3)\phi(x_1)\phi(x_4)  +\Delta(x_2-x_4)\phi(x_1)\phi(x_3) \cr
& +\Delta(x_3-x_4)\phi(x_1)\phi(x_2) + \Delta(x_1-x_2)\Delta(x_3-x_4)+\Delta(x_1-x_3)\Delta(x_2-x_4) \cr
&+\Delta(x_2-x_3)\Delta(x_1-x_4) \ .
\label{nagrandepresaperculo}\end{align}
It is interesting to consider the case of normal ordered potentials. This is the way ${\cal D}^+_{\phi}$ enters in the formulation. Namely, according to
(\ref{babablaxy}) we have
\be
:F[\phi]:=F[{\cal D}^+_{\phi}]\cdot 1 \ .
\ee
Note that ${\cal D}^\pm_{\phi}$ acts as the inverse of ${\cal D}^\mp_{\phi}$
\be
F(\phi)= (F({\cal D}^+_{\phi})\cdot 1|_{\phi={\cal D}^-_{\phi}})\cdot 1 = (F({\cal D}^-_{\phi})\cdot 1|_{\phi={\cal D}^+_{\phi}})\cdot 1 \ .
\ee
Since ${\cal D}^+_\phi$ and ${\cal D}^-_\phi$ differ only by the sign of $\Delta(x-y)$, we can easily get the expression of $\prod_k^n{\cal D}^+_{\phi}(x_k)$ from the one of
$\prod_k^n{\cal D}^-_{\phi}(x_k)$. For example, by (\ref{nagrandepresaperculo}),
\begin{align}
: \phi^2(x) :&= \phi^2(x)-
\Delta(0) \ , \cr
:\phi^3(x): & = \phi^3(x)-3\Delta(0)\phi(x) \ , \cr
:\phi^4(x): & = \phi^4(x)-6\Delta(0)\phi^2(x)+3\Delta^2(0) \ .
\label{oidKl}\end{align}
Another feature of ${\cal D}^{\pm}_{\phi}$ concerns the case of a product of functionals of $\phi$. As shown in (\ref{accordings}), the action
of $\exp\big(\pm{{1\over2}  \delta_{\phi}\Delta\delta_{\phi}}\big)$ on $F[\phi]G[\phi]$ is rather involved. On the other hand, using ${\cal D}^{\pm}_{\phi}$ we have
\be
\exp\big({{1\over2}  \delta_{\phi}\Delta\delta_{\phi}}\big)F[\phi]G[\phi] = F[{\cal D}^-_{\phi}]G[{\cal D}^-_{\phi}]\cdot 1 \ ,
\ee
and
\be
\exp\big(-{{1\over2}  \delta_{\phi}\Delta\delta_{\phi}}\big)F[\phi]G[\phi] = F[{\cal D}^+_{\phi}]G[{\cal D}^+_{\phi}]\cdot 1 \ .
\ee

\noindent
A feature of the dual representation of the generating functional is that it can be expressed in terms of the vev, with respect to the free vacuum, of $:\exp\big(-\int V(\hat\phi)\big):$.
More precisely, we have
\be
T[\phi_c] = N\exp({\cal H}_{\phi_c}) \langle 0 | T:  \exp\Big(-\int V(\hat\phi)\Big):  \exp\Big(\int \hat\phi\Delta^{-1}\phi_c\Big)   | 0\rangle \ ,
\label{sedicigt}\ee
where
\be
{\cal H}_{\phi_c}={1\over2} {\cal D}^-_{\phi_c}\Delta^{-1}{\cal D}^-_{\phi_c} \ ,
\ee
is the ``conjugated free Hamiltonian operator''.
To prove Eq.(\ref{sedicigt}) we first note that
the identification of the
two representations of the generating functional Eq.(\ref{daglie}) can be also expressed in the form
\be
\exp \Big(-\int V(\phi_c)\Big) = \exp\Big(-{1\over2}J\Delta J\Big):\exp\Big({-\int V(\delta_J)}\Big): \exp\Big({1\over2}J\Delta J\Big) \ ,
\label{lhsfg}\ee
obtained by replacing $e^{-\int V}$ in Eq.(\ref{daglie}) by $:e^{-\int V}:$. On the other hand, using the vev representation of the right hand side of (\ref{lhsfg}), we get
\be
\exp \Big(-\int V(\phi_c)\Big)=N\exp\Big(-{1\over2}\phi_c\Delta^{-1}\phi_c\Big)\langle 0 | T:  \exp\Big(-\int V(\hat\phi)\Big):  \exp\Big(\int \hat\phi\Delta^{-1}\phi_c\Big)  | 0\rangle   \ .
\label{foll}\ee
Eq.(\ref{sedicigt}) then follows by the identity
\be
\exp(-U_0[\phi_c])\exp\Big({1\over2}{\delta\over \delta \phi_c} \Delta{\delta\over\delta \phi_c}\Big)\exp(U_0[\phi_c])=\exp({1\over2} {\cal D}^-_{\phi_c}\Delta^{-1}{\cal D}^-_{\phi_c}) \ .
\ee
and then using the definition of $T[\phi_c]$.

\section*{Acknowledgements} It is a pleasure to thank Antonio Bassetto, Kurt Lechner,  Paolo Pasti,
 Dima Sorokin, Roberto Volpato and Jean-Bernard Zuber for interesting comments and
discussions.


\newpage

\begin{thebibliography}{99}

\bibitem{Dirac:1933xn}
  P.~A.~M.~Dirac,
  ``The Lagrangian in quantum mechanics,''
  Phys.\ Z.\ Sowjetunion {\bf 3} (1933) 64.

\bibitem{Feynman} R.~Feynman and R.~Hibbs, Quantum Mechanics and Path Integrals, Dover 2010.

\bibitem{FriedBook} H.~Fried, Functional Methods and Models in Quantum Field Theory, MIT Press, 1972.

\bibitem{Keller:1991bz}
  G.~Keller and C.~Kopper,
  ``Perturbative renormalization of composite operators via flow equations. 1.,''
  Commun.\ Math.\ Phys.\  {\bf 148} (1992) 445.
  doi:10.1007/BF02096544

\bibitem{Salmhofer:1998nk}
  M.~Salmhofer,
  ``Continuous renormalization for fermions and Fermi liquid theory,''
  Commun.\ Math.\ Phys.\  {\bf 194} (1998) 249.
  doi:10.1007/s002200050358

\bibitem{Matone:2015nxa}
  M.~Matone,
  ``Quantum Field Perturbation Theory Revisited,''
  Phys.\ Rev.\ D {\bf 93} (2016) no.6,  065021
  doi:10.1103/PhysRevD.93.065021
  [arXiv:1506.00987 [hep-th]].

\bibitem{Streater:1989vi}
  R.~F.~Streater and A.~S.~Wightman,
  ``PCT, spin and statistics, and all that,''
  Princeton, USA: Princeton Univ. Pr. (2000) 207 p.

\bibitem{Strocchi:2013awa}
  F.~Strocchi,
  An introduction to non-perturbative foundations of quantum field theory,
  Int.\ Ser.\ Monogr.\ Phys.\  {\bf 158} (2013) 368 pp. Oxford University Press.

\bibitem{Deligne:1999qp}
  D. Kazhdan,  ``Introduction to QFT'' in: P.~Deligne, P.~Etingof, D.~S.~Freed, L.~C.~Jeffrey, D.~Kazhdan, J.~W.~Morgan, D.~R.~Morrison and E.~Witten,
  ``Quantum fields and strings: A course for mathematicians. Vol. 1, 2,''
  Providence, USA: AMS (1999) 1-1501

\bibitem{Eckmann:1979vq}
  J.~P.~Eckmann and H.~Epstein,
  ``Time Ordered Products And Schwinger Functions,''
  Commun.\ Math.\ Phys.\  {\bf 64} (1979) 95.
  doi:10.1007/BF01197509



\end{thebibliography}
\end{document}